\documentclass[showpacs,preprintnumbers,amsmath,amssymb]{revtex4}
\usepackage{graphicx,psfrag,amsmath,amssymb,amsfonts,bbm,latexsym,color,dcolumn,bm}

\begin{document}
\title{Computation of Casimir forces for dielectrics or intrinsic semiconductors based on the Boltzmann transport equation}

\author{Diego A. R. Dalvit$^1$ and Steve K. Lamoreaux$^2$}

\address{$^1$ Theoretical Division, MS B213, Los Alamos National Laboratory,
Los Alamos, NM 87545, USA}

\address{$^2$ Yale University, Department of Physics,
P.O. Box 208120, New Haven, CT 06520-8120, USA}

\date{\today}


\begin{abstract}
The interaction between drifting carriers and traveling electromagnetic
waves is considered within the context of the classical Boltzmann transport
equation to compute the Casimir-Lifshitz force between media with
small density of charge carriers, including dielectrics and intrinsic semiconductors. We expand upon our previous work [Phys. Rev. Lett. {\bf 101}, 163203 (2008)]
and derive in some detail the frequency-dependent reflection amplitudes
in this theory and compute the corresponding Casimir free energy
for a parallel plate configuration. We critically discuss the
the issue of verification of the Nernst theorem of thermodynamics in Casimir physics, and explicity show that
our theory satisfies that theorem.
Finally, we show how the theory of drifting carriers connects to previous
computations of Casimir forces using spatial dispersion for the
material boundaries.
\end{abstract}

\pacs{42.50.Ct, 12.20.-m, 78.20.-i}
\vspace{2pc}

\maketitle


\section{Introduction}

Quantum vacuum forces acting between dielectric planar surfaces or between
an atom and a dielectric semi-space were computed long ago by
Lifshitz \cite{Lifshitz} in terms of the complex frequency-dependent dielectric  permittivity $\overline{\epsilon}(\omega)$ of the material boundaries. In this original formulation for ideal dielectrics, 
$\overline{\epsilon}(\omega)$  does not include contributions from current carriers, and as such can be called the ``bare" permittivity. 
Extensions of the Lifshitz theory to media with large free charge carrier density, such as metals or highly doped semiconductors, are typically done by adding a frequency-dependent conduction term, computed from the optical data of the material and extrapolated
to low frequencies by different theoretical models (e.g., a Drude-like term $i 4 \pi \sigma_0/\omega$, where $\sigma_0$
is the dc Drude conductivity). The finite temperature Casimir-Lifshitz
force is extremely sensitive to the optical response of the materials
at low frequency, and therefore different theoretical extrapolations
have resulted in conflicting conclusions about the nature of 
the Casimir force between metals and/or highly doped semiconductors. 

Systems with small density of current carriers, such as insulators
or intrinsic-semiconductors, were recently considered by Pitaevskii \cite{Pitaevskii} and by us \cite{us}. In \cite{Pitaevskii} the thermal Lifshitz force between an atom and a conductor with low
charge density was computed in terms of the Green function formalism,
taking into account the penetration of the static component of the fluctuating EM field into the conductor. This approach is quasi-static,
appropriate for the large distance regime of the thermal Lifshitz 
atom-surface interaction. The relevant (longitudinal) Green function,
expressed in terms of an auxiliary static potential field, can be computed assuming that the gas of carriers in nondegenerate. The static potential satisfies the equation $(\nabla^2 - \kappa^2) \varphi = 0$,
where $\kappa^2 = 4 \pi e^2 n_0 / \overline{\epsilon}_0 k_{\rm B} T$.
Here $-e$ is the electron charge, $\overline{\epsilon}_0$ is the static
bare dielectric constant of the medium (which does not take into 
account the contribution from current carriers), and $n_0$ is the
(uniform) carrier density. Note that $\kappa=1/R_{\rm D}$ is the inverse
of the Debye-H\"uckel screening radius $R_{\rm D}$. For good metals
the Debye radius is very small (on the order of inter-atomic distances),
while for insulators and intrinsic semiconductors it is much larger (on
the order of microns or more). 

In \cite{us} we have extended Pitaevskii's calculation beyond
the quasi-static limit and proposed a theory for the Casimir
interaction taking into account Debye screening and carrier
drift based on the classical Boltzmann equation. Rather than
computing the force with the Green function formalism, we use
the form of the Lifshitz formula written in terms of frequency-dependent
reflection amplitudes $r^p_{{\bf k},j}(w)$ of the $j$-th material
boundary. Here $p$ denotes the polarization of incoming waves
(transverse electric TE or transverse magnetic TM). For simplicity,
we will assume that the material is such that there is no mixing
of polarizations upon reflection (the more general case can be
treated replacing the reflection amplitudes by $2\times 2$ reflection
matrices). The projection on the plane of the interface of the linear momentum of incoming waves is denoted by ${\bf k}$. The Casimir-Lifshitz
pressure between two plane semi-spaces separated by a gap of length $d$
is 
\begin{equation}
P(d) = 2 k_{\rm B} T \sum_{n=0}^{\infty '} \int \frac{d^2 {\bf k}}{(2 \pi)^2} \sqrt{k^2 + \xi^2/c^2} \sum_p \frac{r_1^p r_2^p 
e^{-2 d \sqrt{k^2 + \xi^2/c^2} } }{1 - r_1^p r_2^p 
e^{-2 d \sqrt{k^2 + \xi^2/c^2} } } .
\label{pressure}
\end{equation}
The prime in the sum over $n$ means that the zero frequency $n=0$ term
has to be multiplied by a $1/2$ factor, and all reflection coefficients
are evaluated at imaginary frequencies $\omega=i \xi_n$, where
$\xi_n=2 \pi n k_{\rm B} T/\hbar$ are the Matsubara frequencies.


\section{From Boltzmann transport equation to reflection amplitudes}

In order to compute the appropriate frequency-dependent reflection coefficients
for materials with small density of carriers we will consider that
the EM field interacts with the gas of drifting carriers, and that these
can be modeled as a continuum nondegenerate system. Under these
conditions, it is reasonable to model the carriers with the classical
Boltzmann transport equation coupled to Maxwell's equations for the
electromagnetic field \cite{Sumi,Thiennot}.

For a dielectric the carriers are charged particles (electrons or
ions) hopping from site to site of the crystalline array. For an 
intrinsic semiconductor, the density of carriers and hole is equal,
but their dynamics are different; however, in this work we treat them
as dynamically equivalent, which doubles the charge density. 
Assuming that there is no external applied field on the material, and all fields have a time dependency of the form $e^{-i \omega t}$, Maxwell's
equations take the form
\begin{eqnarray}
\nabla \times {\bf E} = i \mu_0 \omega {\bf H} , \;\; 
& \nabla \times {\bf H} = - i \overline{\epsilon}(\omega) \omega {\bf E} + {\bf J} , \;\; & 
\nabla \cdot {\bf E} = - \frac{e n}{\overline{\epsilon}(\omega)} .
\end{eqnarray}
Here $n$ is the carrier density, $\mu_0$ is the permeability of vacuum,
and ${\bf J}= - e n {\bf v}$ is the carrier current, where ${\bf v}$
is the mean velocity of carriers. The charge
transport in the system is described by the classical Boltzmann
equation,
\begin{equation}
\left( \frac{\partial}{\partial t} + {\bf v} \cdot \nabla \right)
{\bf v} = - \frac{e}{m} {\bf E} - \frac{v^2_T}{n} \nabla n - \frac{{\bf v}}{\tau} , 
\label{Boltzmann}
\end{equation}
where $m$ is the effective mass of charge carriers, $v_T=\sqrt{k_{\rm B} T/m}$ is their mean thermal velocity, and $\tau$ is the carrier
relaxation time. Now we linearize Eq.(\ref{Boltzmann}) with respect to the ac fields. Writing the charge density as $n \rightarrow n_0 + n({\bf r}) e^{-i \omega t}$, the current is ${\bf J} = -e n_0 {\bf v}$ (after
discarding the term $n({\bf r}) {\bf v} e^{-2 i \omega t}$). Since
by assumption there is no external field applied, there is no net
motion of charges, ${\bf v}_0=0$, so the term ${\bf v} \cdot \nabla {\bf v}$ in Eq.(\ref{Boltzmann}) behaves as $e^{-2 i \omega t}$, and can
also be discarded. Finally, the linearized Boltzmann equation is
\begin{equation}
\left( - i \omega  + \frac{1}{\tau} \right) {\bf v} = - \frac{e}{m} 
{\bf E} - \frac{v_T^2}{n_0} \nabla n .
\end{equation}
Inserting $n=- \overline{\epsilon} \nabla \cdot {\bf E} / e$ into
this equation, we can solve for ${\bf v}$:
\begin{equation}
{\bf v}=\frac{\tau}{1-i \omega \tau} \left[ - \frac{e}{m} {\bf E}
+ \frac{v_T^2 \overline{\epsilon}}{e n_0} \nabla \cdot (\nabla \cdot {\bf E} \right].
\end{equation}
Plugging this expression into Maxwell's equations we derive the fundamental equation
for the electric field inside the material,
\begin{equation}
\left[ \nabla^2 + \mu_0 \overline{\epsilon}(\omega) \omega^2 \left(
1+ i \frac{\omega_c}{\omega (1-i \omega \tau)} \right) \right] {\bf E} =
\left[ 1 + i \mu_0 \overline{\epsilon}(\omega)  \frac{\omega D}{1-i \omega \tau} \right]  \nabla \cdot (\nabla \cdot {\bf E}) .
\label{main}
\end{equation}
Here $\omega_c=4 \pi e n_0 \mu / \overline{\omega}$, $\mu=e \tau/m$ is the mobility of carriers, and $D=v^2_T \tau$ is the diffusion constant.
Note that the frequency-dependent ratio $\omega_c/D=4 \pi e^2 n_0/\overline{\epsilon}(\omega) k_{\rm B} T$ coincides with $\kappa^2=1/R^2_{\rm D}$ in the quasi-static limit.

As we will show below, Eq. (\ref{main}) allows TM and TE solutions, so
that there is no cross-polarization upon reflection on the material,
and we can safely use reflection amplitudes rather than reflection matrices in the Lifshitz formula. Let us assume that the material occupies the semi-space region $z<0$ and the region $z>0$ is
vacuum. 

\subsection{TM modes}

For transverse magnetic modes $e_y=0$, so that the electric field
is 
\begin{equation}
{\bf E}({\bf r}) = [ e_x(z) \hat{\bf x} + e_z(z) \hat{\bf z} ] e^{i k x} ,
\end{equation}
where, from now on, we are omitting the phase factors $e^{-i \omega t}$.
Substituting this into Eq. ({\ref{main}) we obtain two coupled second-order differential equations for $e_x$ and $e_z$:
\begin{eqnarray}
&&\left[ \partial^2_z + \mu_0 \overline{\epsilon}(\omega) \omega \left(
1+ i \frac{\tilde{\omega}_c}{\omega} + i \frac{\tilde{D} k^2}{\omega} \right) \right] e_x = i k [ 1 + i \mu_0 \overline{\epsilon}(\omega) \omega \tilde{D}] \partial_z e_z , \nonumber \\
&& \left[ -i \mu_0 \overline{\epsilon}(\omega) \omega \tilde{D} \partial^2_z
- k^2 + \mu_0 \overline{\epsilon}(\omega) \omega^2 \left( 1 + i 
\frac{\tilde{\omega}_c}{\omega} \right) \right] e_z  =
i k [ 1+ i \mu_0 \overline{\epsilon}(\omega) \omega \tilde{D} ] 
\partial_z e_x , \nonumber
\end{eqnarray}
where $\tilde{\omega}_c \equiv \omega_c / (1-i \omega \tau)$ and
$\tilde{D} \equiv D / (1-i \omega \tau)$. 
It is possible to combine these two coupled equations 
into two uncoupled fourth-order differential equations.
To this end one takes the $\partial^2_z$ derivative of the first
equation above, which results in terms proportional to 
$\partial^4_z e_x$, $\partial^2_z e_x$, and $\partial^3_z e_z$.
This last term $\partial^3_z e_z$  can be obtained from taking the $\partial_z$ derivative of the second equation above, which results in terms proportional to $\partial^2_z e_x$ and $e_x$. Putting all together,
one can derive the following fourth-order differential equation
for $e_x$:
\begin{equation}
(\partial_z^2 - \eta^2_T) \; (\partial_z^2 - \eta^2_L) e_x = 0 ,
\end{equation}
where
\begin{eqnarray}
\eta^2_T(\omega) = k^2 - \mu_0 \overline{\epsilon}(\omega) \omega^2 \left(
1+ i \frac{ \tilde{\omega}_c}{\omega} \right) , \\
\eta^2_L(\omega) = k^2 - i \frac{\omega}{\tilde{D}} \left( 1+ i \frac{
\tilde{\omega}}{\omega} \right).
\end{eqnarray}
In a similar fashion one obtains the following equation for $e_z$:
\begin{equation}
(\partial_z^2 - \eta^2_T) \; (\partial_z^2 - \eta^2_L) e_z =0.
\end{equation}
The solutions of these equations that vanish for $z \rightarrow - \infty$ are $e_x(z)=A_T e^{\eta_T z} + A_L e^{\eta_L z}$ and
$e_z(z)=A'_T e^{\eta_T z} + A'_L e^{\eta_L z}$, where we assume
${\rm Re} \eta_T>0$ and ${\rm Re} \eta_L>0$. The amplitudes
are related as $A'_L=-i \eta_L A_L / k$ and $A'_T=-i k A_T / \eta_T$.
Given this TM electric field, the associated TM magnetic field can be readily
computed, ${\bf H}=i \hat{\bf y} A_T e^{\eta_T z} e^{i k x} (k^2-\eta^2_T) / \mu_0 \omega \eta_T$.

Now we compute the TM reflection amplitude, imposing the boundary
conditions on the $z=0$ interface. These conditions are 
$E_x$, $H_y$, and $\overline{\epsilon} E_z$ continuous (the continuity
of $B_z$ is automatically satisfied for TM modes). On the vacuum side
($z>0$) the condition $\nabla \cdot {\bf E}=0$ implies $i k e_x + \partial_z e_z =0$, so that the fields incident on the interface from
the $z>0$ side are
${\bf E}^{\rm in} = E_0 [ - (k_z/k) \hat{\bf x} + \hat{\bf z} ] e^{i k_z z} e^{i k x}$ 
and
${\bf H}^{\rm in} = - E_0 \omega  \hat{\bf y} e^{i k_z z} e^{i k x} /
\mu_0 k c^2$,
where we have used that in vacuum $k^2+k_z^2=\omega^2/c^2$. The reflected fields are
${\bf E}^{\rm r} = r E_0 [ + (k_z/k) \hat{\bf x} + \hat{\bf z} ] e^{- i k_z z} e^{i k x}$ 
and
${\bf H}^{\rm r} = - r E_0 \omega  \hat{\bf y} e^{-i k_z z} e^{i k x} /
\mu_0 k c^2$,}
where $r$ is the reflection amplitude. The transmitted fields into the
material ($z<0$) are
${\bf E}^{\rm t} =
\left[ \left( A_T e^{\eta_T z} + A_L e^{\eta_L z} \right) \hat{\bf x} + 
\left( - \frac{i k}{\eta_T} A_T e^{\eta_T z} - \frac{i \eta_L}{k} A_L
e^{\eta_L z} \right) \hat{\bf z} \right] e^{i k x}$ 
and
${\bf H}^{\rm t} =
i \hat{\bf y}
(k^2- \eta^2_T) A_T e^{\eta_T z} e^{i k x} / \mu_0 \omega \eta_T$.
Imposing the boundary conditions, and after some straightforward algebra,
the reflection amplitude can be written as $r(\omega)=(1-\alpha)/(1+\alpha)$, where $\alpha=\frac{k^2}{i \eta_L k_z} \left[ 
\frac{1}{\overline{\epsilon}(\omega)} - \frac{\omega^2/c^2}{k^2-\eta^2_T}
+ \frac{\eta_L \eta_T \omega^2/c^2}{k^2 (k^2-\eta^2_T)} \right]$. Expressed along imaginary frequencies $\omega=i \xi$, the TM reflection
amplitude is
\begin{equation}
r^{\rm TM}_{\bf k}(i \xi) = \frac{\overline{\epsilon}(i \xi) \sqrt{k^2 + \xi^2/c^2} - \chi}{\overline{\epsilon}(i \xi) \sqrt{k^2 + \xi^2/c^2} + \chi}, 
\label{rTM}
\end{equation}
where 
\begin{equation}
\chi=\frac{1}{\eta_L} \left[ k^2 + \overline{\epsilon}(i \xi) 
\frac{\xi^2}{c^2} \frac{\eta_L \eta_T -k^2}{\eta^2_T - k^2} \right] .
\end{equation}
Along imaginary frequencies, $\eta_L$ and $\eta_T$ take the form:
\begin{eqnarray}
&& \eta_L(i \xi) = \sqrt{
k^2+ \frac{4 \pi e^2 n_0}{\overline{\epsilon}(i \xi) k_{\rm B} T} + 
\frac{\xi (1+ \xi \tau)}{v^2_T \tau} 
} , 
\label{etaL} \\
&& \eta_T(i \xi) = \sqrt{
k^2 + \overline{\epsilon}(i \xi) \frac{\xi^2}{c^2} 
\left( 
1+ \frac{4 \pi e^2 n_0 \tau}{m \overline{\epsilon}(i \xi) \xi (1+ \xi \tau)}
\right) }  =
\sqrt{k^2 + [\overline{\epsilon}(i\xi) + 4 \pi \sigma(i\xi)/\xi] \xi^2/c^2 } ,
\label{etaT}
\end{eqnarray}
where $\sigma(i\xi)=\sigma_0/(1+\xi \tau)$ and 
$\sigma_0=e^2 n_0 \tau/m$ are the ac and dc Drude conductivities,
respectively. Therefore, Eq. (\ref{rTM}) gives a modified Fresnel
TM coefficient due to the presence of Debye-H\"uckel screening 
and charge drift in the material.

\subsection{TE modes}

For transverse electric modes $e_z=0$, so that the electric field
is
\begin{equation}
{\bf E}({\bf r})=[e_x(z) \hat{\bf x} + e_y(z) \hat{\bf y} ] e^{i k x} .
\end{equation}
Substituting this into Eq. ({\ref{main}) we obtain two second-order differential equations for $e_x$ and $e_z$:
\begin{eqnarray}
&& 
\left[ \partial^2_z + \mu_0 \overline{\epsilon}(\omega) \omega^2 
\left( 1+ i \frac{\tilde{\omega}_c}{\omega} +
i \frac{\tilde{D} k^2}{\omega} \right) \right] e_x =0 , \nonumber \\
&&
\left[ \partial^2_z - k^2 + \mu_0 \overline{\epsilon}(\omega) \omega^2
\left(1+ i \frac{\tilde{\omega}_c}{\omega} \right) \right] e_y =
i k [ 1 + i \mu_0 \overline{\epsilon}(\omega) \omega \tilde{D} ]
\partial_z e_x .
\nonumber
\end{eqnarray}
The solution to the first equation is $e_x(z)=A e^{\beta z}$, 
where $A$ is a constant and $\beta^2=-i \mu_0
\overline{\epsilon}(\omega) \omega \tilde{D} \eta^2_L$
(we assume ${\rm Re} \beta>0$). Plugging this solution into the second
equation we obtain $e_y(z)=B e^{\eta_T z} +
C e^{\beta z}$, where 
$C=i k A \beta (1+i \mu_0 \overline{\epsilon}(\omega) \omega \tilde{D})/(\beta^2-\eta^2_T)$ . Given this
TE electric field, the associated TE magnetic field is
${\bf H}=(1/i \mu_0 \omega) [-(B \eta_T e^{\eta_T z} + C \beta e^{\beta z}) \hat{\bf x} + A \beta e^{\beta z} \hat{\bf y} +
i k (B e^{\eta_T z} + C e^{\beta z}) \hat{\bf z} ] e^{i k x}$.

Now we compute the TE reflection amplitude imposing the boundary
conditions on the interface (the continuity of $\overline{\epsilon} E_z$ is automatically satisfied for TE modes). On the vacuum side we 
have $e_x=0$, so that the incident fields are
${\bf E}^{\rm in}=E_0 e^{i k_z z} e^{i k x} \hat{\bf y}$ and
${\bf H}^{\rm in}=E_0 (-k_z \hat{\bf x} + k \hat{\bf z}) e^{i k_z z} e^{i k x} / \mu_0 \omega$, and the reflected fields are
${\bf E}^{\rm r}=r E_0 e^{-i k_z z} e^{i k x} \hat{\bf y}$ and
${\bf H}^{\rm r}=r E_0 (+k_z \hat{\bf x} + k \hat{\bf z}) e^{-i k_z z} e^{i k x} / \mu_0 \omega$. The transmitted fields into the material
are given above. A simple calculation leads to the expression 
of the reflection amplitude $r=(i k_z - \eta_T)/(i k_z + \eta_T)$.
Upon performing the rotation $\omega \rightarrow i \xi$, we get
\begin{equation}
r^{\rm TE}_{\bf k}(i \xi) = \frac{ \sqrt{k^2+\xi^2/c^2} - \eta_T}
{ \sqrt{k^2+\xi^2/c^2} + \eta_T}.
\label{rTE}
\end{equation}
Using Eq. (\ref{etaT}) we see that $r^{\rm TE}$
is the usual Fresnel TE reflection coefficient with a dielectric
permittivity equal to the sum of the ``bare" one and the ac Drude 
(conduction) permittivity, $\epsilon(i \xi)= \overline{\epsilon}(i\xi) + 4 \pi \sigma(i\xi)/\xi$.

As discussed in detail in \cite{us}, these modified TE and TM reflection coefficients have appropriate limiting behaviors. In the quasi-static limit ($\xi \rightarrow 0$) they coincide with the ones derived in \cite{Pitaevskii} for conductors with small density of carriers in the large distance (low frequency) regime, namely $r^{\rm TE}_{\bf k}(0)=0$ (in the
static limit the TE polarized field is a pure magnetic field, which fully penetrates the nonmagnetic material) and $r^{\rm TM}_{\bf k}(0)=(\overline{\epsilon}_0 q-k)/(\overline{\epsilon}_0 q+k)$, with $q=\sqrt{k^2+\kappa^2}$ (in the static limit $r^{\rm TM}$ interpolates between a good conductor and an ideal dielectric). On the other hand, for any frequency $\xi \ge 0$, and in the limit of ideal dielectrics (small free charge density and small
effective thermal velocity), we recover the usual Fresnel
equations written in terms of the bare permittivity $\overline{\epsilon}(\omega)$. 


\section{Influence of drifting carriers in the Casimir-Lifshitz free energy}

We now study the implications of our theory in the computation
of the Casimir-Lifshitz free energy 
\begin{equation}
\frac{E}{A}= k_{\rm B} T \sum_p \sum_{n=0}^{\infty '} 
\int \frac{d^2{\bf k}}{(2 \pi)^2} 
\ln[ 1-r^p_{{\bf k},1}(i \xi_n) r^p_{{\bf k},2}(i \xi_n) e^{-2 d \sqrt{k^2+\xi_n^2/c^2}} ] ,
\label{freeenergy}
\end{equation} 
between two identical planar
semi-spaces with small density of charge carriers, such as intrinsic
semiconductor media. As examples, we consider the cases of pure germanium and pure silicon. The reflection coefficients Eqs.(\ref{rTM},\ref{rTE}) that enter into this equation depend on temperature explicitly through the Matsubara frequencies and implicitly through
the optical and conductivity parameters, which we proceed to quote.

For intrinsic Ge, the bare permittivity can be approximately fitted with a
Sellmeier-type expression 
\begin{equation}
\overline{\epsilon}(i \xi) = \overline{\epsilon}_{\infty} +
\omega^2_0 \frac{\overline{\epsilon}_0 - \overline{\epsilon}_{\infty}}{\xi^2 + \omega_0^2} ,
\end{equation}
where $\overline{\epsilon}_0 \approx 16.2$, $\overline{\epsilon}_{\infty} \approx 1.1$, and $\omega_0 \approx 5.0 \times 10^{15}$rad/sec at $T \approx 300$K. The temperature dependence of the permittivity has been measured in the $20-300$K range at wavelengths $1.9-5.5 \mu$m \cite{Frey}, and shown to be very weak. Therefore, in this paper we assume that the permittivity is approximately constant as a function of temperature, and given by the above Sellmeier fitting function. 
The intrinsic carrier density varies with temperature as
\begin{equation}
n_0(T) = \sqrt{n_c n_v} e^{-\frac{E_g}{2 k_{\rm B} T}} ,
\end{equation}
where $n_c$ and $n_v$ are the effective density of states
in the conduction and valence band, respectively. These depend on temperature as $n_c(T)=1.98 \times 10^{15} T^{3/2} {\rm cm}^{-3}$ and $n_v(T)=9.6 \times 10^{14} T^{3/2} {\rm cm}^{-3}$ (temperature is measured in degrees K). The band gap energy also depends on temperature, $E_g(T)=0.742 - 4.8 \times 10^{-4} T^2/(T+235)$eV. 
The effective mass of conductivity is $m=0.12 m_e$, where $m_e$
is the free electron mass \cite{parameters}. The relaxation time
depends on temperature as $\tau(T)= \tau_0 + \tau_1 e^{C_1 (T/300)^2 + C_2 (T/300)}$
\cite{relaxation}, where $\tau_0=0.26$ps, $\tau_1=1.49$ps, $C_1=-0.434$, and $C_2=1.322$.
At $T=300$K one has $E_g=0.66$eV, $n_c=1.0 \times 10^{19} {\rm cm}^{-3}$, $n_v=5.0 \times 10^{18} {\rm cm}^{-3}$, and $\tau=3.9$ps.

For intrinsic Si, $\overline{\epsilon}_0 \approx 11.87$, $\overline{\epsilon}_{\infty} \approx 1.035$, and $\omega_0 \approx 6.6 \times 10^{15}$rad/sec at $T \approx 300$K. The temperature dependence of the permittivity has been measured in the $20-300$K range at wavelengths $1.1-5.6 \mu$m \cite{Frey}, and also shown to be very weak. The effective density
of states in the conduction and valence bands are
$n_c(T)=6.2 \times 10^{15} T^{3/2} {\rm cm}^{-3}$ and $n_v(T)=3.5 \times 10^{15} T^{3/2} {\rm cm}^{-3}$ respectively.
The band gap energy is 
$E_g(T)=1.17 - 4.73 \times 10^{-4} T^2/(T+636)$eV. 
The effective mass of conductivity is $m=0.26 m_e$ \cite{parameters}. The relaxation time parameters are 
$\tau_0=1.0$ps, $\tau_1=-0.538$ps, $C_1=0.0015$, and $C_2=-0.09$
\cite{relaxation}.
At $T=300$K one has $E_g=1.12$eV, $n_c=3.2 \times 10^{19} {\rm cm}^{-3}$, $n_v=1.8 \times 10^{19} {\rm cm}^{-3}$, and $\tau=0.5$ps.

In Fig 1 we plot the Casimir-Lifshitz free energy between two identical
planar intrinsic semiconducting (Ge and Si) semi-spaces as a function of the distance between them. We use our theory of Casimir forces
with account of Debye-H\"uckel screening and charge drift to compute
the reflection coefficient Eqs. (\ref{rTM},\ref{rTE}) and compare
these predictions with the simple model in which the reflection
coefficients are given by the usual Fresnel formulas in which
the permittivity of the materials $\epsilon(i\xi)$ is computed
by adding to the bare permittivity $\overline{\epsilon}(i\xi)$
a dc conductivity term $4 \pi \sigma_0/\xi$. In both models we normalize the free energies to the free energy computed using the
standard Lifshitz theory using the bare permittivity only.
For intrinsic Ge and intrinsic Si, $\tilde{\omega}_c$ and
$\tilde{D}/\xi$ are both very small in the relevant range of frequencies for the Lifshitz formula. Therefore only
the $n=0$ TM mode is significantly modified by the screening
and charge drift effects, $r^{\rm TE}_{\bf k}(0)=0$ and
$r^{\rm TM}_{\bf k}(0)=(\overline{\epsilon}(0) q - k)/
(\overline{\epsilon}(0)q + k)$, with $q=\sqrt{k^2+\kappa^2}$,
as in \cite{Pitaevskii}. In all other $n \ge 1$ terms in 
Eq. (\ref{freeenergy}) the reflection coefficients can be
replaced by the standard Fresnel expressions in terms of the
bare permittivity $\overline{\epsilon}(i \xi)$. Since for intrinsic 
carrier density for Ge ($\approx 10^{13} {\rm cm}^{-3}$) is much larger than
that for Si ($\approx 10^{10} {\rm cm}^{-3}$), the Debye radius is of
Ge, $R_{\rm D}=1/\kappa=0.68 \mu$m is much smaller than that of
Si, $R_{\rm D}=24 \mu$m. As follows from Fig. 1, the effect of Debye screening and drifting  carriers (denoted as ``drift" in Fig. 1) becomes important for distances much larger than the Debye
radius, so that this effect is more
likely to be detected in Ge than in Si. In the latter case, for distances $d>R_{\rm D}$, the Casimir force is too weak, 
at such a large distance, to be measured by any current or proposed experimental technique.
From Fig. 1 we also
note that when $d \gg R_{\rm D}$ the plates appear as perfect
conductors for the TM $n=0$ mode, while in the case of the additive
term ($\overline{\epsilon}(i\xi)+4 \pi \sigma_0/\xi$, denoted as ``cond" in the figure), the plates appear as perfect conductors for the TM $n=0$ mode at distances of the order of $\lambda_T=\hbar c/k_{\rm B} T$ ($\approx 7 \mu$m at $T=300$K), independent of the material properties.

\begin{figure}[t]
\begin{center}
\includegraphics[width=0.5\columnwidth]{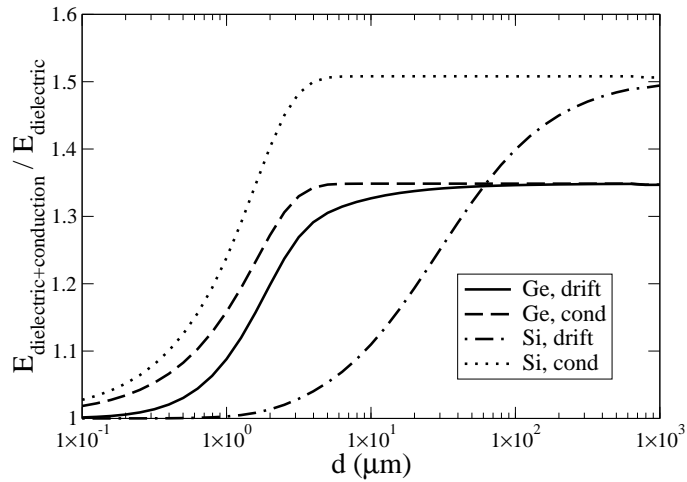}
\end{center}
\caption{Casimir-Lifshitz free energy at $T=300$K for intrinsic
Ge and Si taking into account charge drift Eqs. (\ref{rTM},\ref{rTE})
(curves denoted by ``drift") or an additive dc conductivity term 
$4 \pi \sigma_0/\xi$ to the bare permittivity (curves denoted by ``cond"). 
For Ge, $\sigma_0=1/(43 \Omega \; {\rm cm})$, and for Si, $\sigma_0=1/(2.3 \times 10^5 \Omega \; {\rm cm})$. The free energies are normalized to those computed using only the bare permittivity in the usual Lifshitz theory.}
\label{Fig1}
\end{figure}


\section{On the satisfaction of Nernst theorem as a prerequisite 
for a Casimir theory}

Calculations of finite temperature Casimir-Lifshitz forces between media with large charge density, such as metals and highly doped semiconductors, have resulted in a heated debate on the adequate way to describe the optical properties of such systems within the Lifshitz formalism. Different phenomenological ways to extrapolate optical data to low frequencies, either with a Drude model including dissipation or
with a plasma model setting dissipation to zero from the start, result
in completely different predictions for the force at finite temperature \cite{controversy}.

It has been suggested that the Nernst theorem of 
thermodynamics serves as a way to accept or discard conductivity
models when applied to the computation of the Casimir-Lishitz entropy
$S(T)=- \partial E(T) /\partial T$. The Nernst theorem
states that the entropy of a physical system of $N$ particles in thermal equilibrium at zero temperature is a well-defined constant, determined only by the degeneracy $\Omega_N$ of the ground state of the system, that is, 
$S(T=0)=k_{\rm B} \ln \Omega_N$. For systems with non-degenerate ground states $\Omega_N=1$ (e.g., a perfect crystal lattice), the entropy should vanish at zero temperature. This is not the case for a large class of systems, including spin networks and glasses, that can have a large collection of degenerate ground states (with degeneracy
$\Omega_N$ depending on the total number of particles),
so that $S(T=0)>0$ in such systems. 
In some textbooks \cite{Rumer} it is further required
as part of Nernst theorem that the degeneracy $\Omega_N$ be independent of any varying parameters of the system (such as pressure, volume, field intensities, etc). This has been used by some authors to discard Casimir theories which lead to a zero-temperature entropy that depends on the separation $d$ between the Casimir plates. The requirement of independence of $S(T=0)$ on volume is at odds with the fact that entropy is an extensive quantity, and should grow with system size \cite{correctNernst}. It is not clear to us
that one can simply discard a model of conductivity
for the Casimir plates based on the fact that the zero
temperature entropy depends on the distance between plates. 
We believe that this issue requires further study.
Of course, if for a given theoretical model for the 
Casimir plates $S(T=0)<0$, then such model violates Nernst theorem.

In the remainder of this section we explicitly prove that
our theory for Casimir forces with intrinsic semiconductor
media is compatible with Nernst theorem of thermodynamics,
resulting in $S(T=0)=0$, as for systems with a non-degenerate
ground state ($\Omega_N=1$). We will closely follow the
approach in \cite{Intravaia}. We start by expressing the Casimir free energy as $E = \frac{\hbar}{2\pi} \sum_{p,{\bf k}} \sum_{n=0}^{\infty '} \theta g^p(i n \theta,{\bf k};\theta)$, where
\begin{equation}
g^p(\omega,{\bf k};\theta)=
\ln[1-r^p_{{\bf k},1}(\omega,\theta) \;  r^p_{{\bf k},2}(\omega, \theta) e^{-2 d \sqrt{k^2 - \omega^2/c^2}} ],
\end{equation}
and $\theta=2 \pi k_{\rm B} T/\hbar$. Note that we have
allowed for an implicit dependence of the reflection coefficients on temperature. The Casimir-Lifshitz
entropy is $S=-(2 \pi/\hbar) \partial E/\partial \theta$. In Fig. 2
we plot the behaviour of $g^p(i\xi,{\bf k};\theta)$ as a function of the
imaginary frequency $\omega=i \xi$ and as a function of 
$k=|{\bf k}|$ for TM and TE polarizations for different temperatures (the corresponding reflection amplitudes
are obtained from Eqs. ({\ref{rTM},\ref{rTE})). Let us 
consider the TE and TM contributions to the entropy
separately. 

\begin{figure}[t]
\hspace{-10pt}
\scalebox{2.0}{\includegraphics{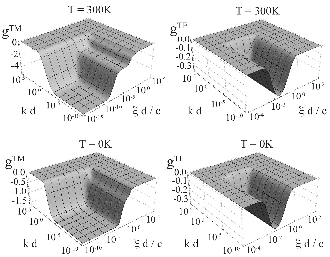}}

\caption{Behavior of the functions $g^p(i\xi,{\bf k})$ used to compute the Casimir-Lifshitz free energy and entropy for semiconductor materials with account of drifting carriers.
The reflections coefficients are given by (\ref{rTM}) and (\ref{rTE}),
parameters are for intrinsic Ge (see text), and the
distance is set to $d=1\mu$m. The variation with temperature
(in the range $T=0-300$K) of the TE function is not perceptible on the scale of the figure. The
corresponding functions without account of Debye screening and carrier drift correspond to the $T=0$K plots in this figure.}
\end{figure}

For TE modes, since the reflection coefficient (\ref{rTE})
depends implicitly on temperature only through $\tilde{\omega}_c$, which becomes exponentially small as low temperatures because the carrier density vanishes
as $T \rightarrow 0$, it is possible to show that the
$\theta \rightarrow 0$ and $\omega \rightarrow 0$ limits
of $r^{\rm TE}_{{\bf k}}(\omega,\theta)$ commute,
that $g^p(\omega,{\bf k}; \theta)$ is analytic in the upper-half complex 
$\omega$ plane, and that the sum over $n$ and the derivative with respect to $\theta$ in the expression for the entropy can be interchanged. Therefore, the contribution of TE modes to the entropy is
\cite{Intravaia}
\begin{equation}
S^{\rm TE}(T) = - \sum_{\bf k} \sum_{n=0}^{\infty '} [ 
g^{\rm TE}(i n \theta,{\bf k}; \theta) + i n \theta g^{\rm TE}_{\omega}(i n \theta,{\bf k}; \theta) + \theta g^{\rm TE}_{\theta}(i n \theta, {\bf k}; \theta) ] .
\label{entropytheory}
\end{equation}
Here we have defined $g^{\rm TE}_{\omega}\equiv \partial_{\omega}
g^{\rm TE}(\omega,{\bf k}; \theta)$ and 
$g^{\rm TM}_{\theta} \equiv \partial_{\theta}
g^{\rm TM}(\omega,{\bf k};\theta)$. Using the analytical properties
of the function $g^{\rm TE}$ it is possible to write
an expansion of the first two terms in (\ref{entropytheory}) 
in powers of temperature, resulting in 
$(g^{\rm TE}_{\xi}(0,{\bf k})/6) \tau + (5 g^{\rm TE}_{\xi^2}(0,{\bf k})/12) \tau^2 +\ldots$,
where $g^{\rm TE}_{\xi}(0,{\bf k})=\lim_{\xi\rightarrow 0} \partial g^{\rm TE}(i\xi,{\bf k};0)/\partial \xi$ and $g^{\rm TE}_{\xi^2}(0,{\bf k})=\lim_{\xi\rightarrow 0} \partial^2 g(i \xi,{\bf k};0)/\partial \xi^2$ \cite{Intravaia}. Thus, the first two terms in  (\ref{entropytheory}) give a vanishing entropy at $T=0$, and imply a low-temperature behaviour of the entropy proportional to $T^2$, since $g^{\rm TE}_{\xi}(0,{\bf k})=0$ and
$g^{\rm TE}_{\xi^2}(0,{\bf k}) = - e^{-2 k d} \overline{\epsilon}^2_0 \omega_c^2/ 8 k^4 <0$ (see TE plots in Fig. 2). The last term in
(\ref{entropytheory}) is proportional to $\partial_{\theta} \tilde{\omega}_c$, which is exponentially small at low temperatures. Therefore, the full TE contribution to the entropy vanishes at zero temperature,
namely $S^{\rm TE}(0)=0$.

For TM modes, the reflection coefficient (\ref{rTM}) 
depends implicitly on temperature both through $\tilde{\omega}_c$ and $\tilde{D}$ in a complicated fashion. 
Contrary to the TE case, the $\theta \rightarrow 0$
and $\omega \rightarrow 0$ limits of $r^{\rm TM}(\omega,{\bf k};\theta)$ do not commute, and therefore it is not
possible to write the contribution of TM modes to the entropy
in the simple form (\ref{entropytheory}). The $n\ge 1$ and $n=0$ terms
have to be treated separately. 
This can be done by defining
a new function $\tilde{g}^{\rm TM}(i n \theta, {\bf k})$ which is identical to
$g^{\rm TM}(i n \theta,{\bf k}; \theta)$ for $n \ge 1$, and for
$n=0$ it is defined as $\tilde{g}^{\rm TM}(0,{\bf k}) \equiv
\lim_{\theta \rightarrow 0} g^{\rm TM}(\alpha \theta, {\bf k};\theta)$,
where we approach zero along the path $\omega=\alpha \theta$
\cite{Intravaia}. The TM contribution to the entropy is
\begin{equation}
S^{\rm TM}(T) = - \sum_{\bf k} \left\{ \frac{g^{\rm TM}(0,{\bf k};\theta) - \tilde{g}^{\rm TM}(0,{\bf k})}{2} \right.
\left. + \sum_{n=0}^{\infty'} 
[ \tilde{g}^{\rm TM}(i n \theta, {\bf k}; \theta) +
i n \theta \tilde{g}^{\rm TM}_{\omega}(i n \theta, {\bf k}; \theta) +
\theta \tilde{g}^{\rm TM}_{\theta}(i n \theta, {\bf k}; \theta) ]
\right\}.
\label{entropyTM}
\end{equation}
As in the TE case, the first two terms in the sum $\sum_{n=0}^{\infty '}$
vanish in the zero temperature limit $\theta \rightarrow 0$. Their
first non-vanishing contribution to the entropy is linear in $T$, since
$g^{\rm TM}_{\xi}(0,{\bf k}) > 0$ (see plots TM in Fig. 2). On the other
hand, in the low-temperature limit the third term in the sum over $n$ in (\ref{entropyTM}) is, 
\begin{equation}
\sum_{n=0}^{\infty '} \theta \tilde{g}^{\rm TM}_{\theta}(i n \theta, {\bf k}; \theta) \stackrel{\theta \rightarrow 0}{\longrightarrow} \partial_{\theta} \omega_c(\theta) \int_{0}^{\infty}
d\xi \frac{\partial \tilde{g}^{\rm TM}(i\xi, {\bf k};\theta)}{\partial \omega_c} + \partial_{\theta}D(\theta) \int_{0}^{\infty}
d\xi \frac{\partial \tilde{g}^{\rm TM}(i\xi, {\bf k};\theta)}{\partial D} .
\label{tricky}
\end{equation}
The first term in (\ref{tricky}) is zero at $T=0$ due to the exponential decay of $\omega_c$
at low temperatures. Although $\partial_{\theta} D$ does not vanish at 
$T=0$ (since $\tau(T)$ goes to a non-zero constant at $T=0$ and
then $D(T) \propto T$ at low temperatures), the second term 
in (\ref{tricky}) is also zero at $T=0$ because the integrand is exponentially small. Finally, in the limit $\theta \rightarrow 0$,
the first line in (\ref{entropyTM}) vanishes since
$g^{\rm TM}(0,{\bf k};0) = \tilde{g}^{\rm TM}(0,{\bf k})$ by definition.
Therefore, the TM contribution to the zero temperature entropy is
$S^{\rm TM}(T=0)=0$. We conclude that our theory for Casimir-Lifshitz
forces in systems with low density of carriers (intrinsic semiconductors,
dielectrics, etc) with account of Debye screening and charge drift is in
agreement with Nernst theorem.


\section{Connection between this theory of drifting carriers and
spatial dispersion}

As we have already mentioned, the static limit of the reflection coefficients 
(\ref{rTM},\ref{rTE}) obtained by us using the Boltmann transport approach 
coincide with those previously derived in \cite{Pitaevskii}, where it was noted that the same static reflection amplitudes can be interpreted in terms of spatial dispersion. Indeed, using as the static permittivity
tensor ${\boldsymbol \epsilon}={\rm diag}(\epsilon^{\perp},
\epsilon^{\perp}, \epsilon^{\|})$ with the 
transverse permittivity $\epsilon^{\perp}=\overline{\epsilon}_0$, and
the longitudinal permittivity
depending on wavevector as $\epsilon^{\|}(k)=\overline{\epsilon}_0 [1 + 1/(k R_{\rm D})^2]$, and
computing the reflection coefficients for an anisotropic (uniaxial)
material \cite{MMlongPRA}, it is
straightforward to recover the static versions of (\ref{rTM},\ref{rTE}).

Although the original Lifshitz paper \cite{Lifshitz} for dispersion forces between bodies separated by vacuum did not allow for spatial dispersion, it can certainly be generalized
to include nonlocal dielectric response in those setups,
for example by averaging the vacuum Maxwell stress tensor
and calculating field strengths via the retarded Green
tensor of the field, that should include spatial dispersion 
when it is important \cite{Pitaevskii,Barash}. As noted
in \cite{Pitaevskii}, problems arise when the bodies are
separated by a liquid instead of vacuum.

Instead of proceeding via the Green function method, here we use the scattering formalism generalized to spatial
dispersion to compute the force between plates separated by 
vacuum. The Casimir pressure between the plates is given by the same Eq. (\ref{pressure}), and the effects of spatial dispersion are incorporated by appropriately writing the reflection amplitudes $r^p_{\bf k}(\omega)$ in terms of the permittivity tensor 
${\boldsymbol \epsilon}(\omega,k)$ 
\cite{Sernelius,Esquivel}. In \cite{Sernelius} the permittivity tensor 
is computed in the random phase approximation (Lindhard dielectric
function) including dissipation \cite{Kliewer}. The reflection
amplitudes are written as
\begin{equation}
r^p_{{\bf k}}(\omega) = \frac{H^p(k,\omega) -1}{H^p(k,\omega)+1} ,
\end{equation}
where the TM and TE $H$-functions are (our $H, h$ functions correspond to the $G, g$ functions in \cite{Sernelius})
\begin{eqnarray}
H^{\rm TM}(k,\omega) &=& \frac{k}{\gamma^0} \tilde{h}_a(k,\omega) - 
\frac{(\omega/c)^2}{(\gamma^0)^2} \tilde{h}_b(k,\omega) +
\frac{k (k-\gamma^0)}{(\gamma^0)^2} \tilde{h}_c(k,\omega) + 1 , \\
H^{\rm TE}(k,\omega) &=& \tilde{h}(k,\omega) + 1 ,
\end{eqnarray}
where
\begin{eqnarray}
h_a(k,\omega) &=& 2 k \int_{-\infty}^{\infty} \frac{dq_z}{2 \pi} \frac{1}{q^2 \epsilon^{\|}(q,\omega)} , \nonumber \\
h_b(k,\omega) &=& 2 \gamma^0(k,\omega)\int_{-\infty}^{\infty} \frac{dq_z}{2 \pi} \frac{1}{q^2 - \epsilon^{\perp}(q,\omega) (\omega/c)^2 } , \nonumber \\
h_c(k,\omega) &=& \frac{2(\omega/c)^2 k \gamma^0(k,\omega)}{k-\gamma^0(k,\omega)} \int_{-\infty}^{\infty} \frac{dq_z}{2 \pi} \frac{1}{q^2 
[q^2 - \epsilon^{\perp}(q,\omega) (\omega/c)^2]} .
\label{integrals}
\end{eqnarray}
Here $\gamma^0(k,\omega)=\sqrt{k^2-(\omega/c)^2}$, ${\bf q}=({\bf k},q_z)$, and the tilde above an $h$-function means the $h$-function minus the same function except that the dielectric function is set to unity \cite{Sernelius}. In the limit of neglegible spatial dispersion ($\epsilon^{\|}$ and $\epsilon^{\perp}$ independent of $q$) one gets the usual Fresnel expressions for the reflection coefficients.

We now show that our reflection coefficients can be linked to spatial dispersion even beyond the static limit, connecting in this way our approach based on Boltzmann transport equation with spatial nonlocality. 
We closely follow the approach of \cite{Sernelius}. As in the static case,
we assume that the nonlocal permittivity depends solely on $k$, being
independent of $q_z$. This allows us to straightfowardly compute the integrals in (\ref{integrals}). From the $H^{\rm TE}$ function we obtain
\begin{equation}
\frac{H^{\rm TE}(k,i\xi) -1}{H^{\rm TE}(k,i\xi)+1} = 
\frac{\sqrt{k^2 + \xi^2/c^2} - \sqrt{k^2+ \epsilon^{\perp}(k,i\xi) \xi^2/c^2}}{\sqrt{k^2 + \xi^2/c^2} + \sqrt{k^2+ \epsilon^{\perp}(k,i\xi) \xi^2/c^2}} .
\end{equation}
We recover our TE reflection coefficient (\ref{rTE}) for a transverse
dielectric function independent of $k$,
\begin{equation}
\epsilon^{\perp}(k,i \xi) = \overline{\epsilon}(i \xi) 
\left[ 1 + \frac{\omega_c}{\xi (1 + \xi \tau)} \right] .
\label{transversal}
\end{equation}
For $H^{\rm TM}$ we obtain
\begin{equation}
H^{\rm TM}(k,i\xi) = 1+ \frac{k}{\gamma^0} \left[ \frac{1}{\epsilon^{\|}(k,\omega)} - 1 \right] + 
\frac{\xi^2/c^2}{\gamma^0} \left[ \frac{1}{\eta_T} - \frac{1}{\gamma^0} \right] 
- \frac{k \xi^2/c^2}{\gamma^0} \left[ \frac{1}{k \eta_T - \eta_T^2} - \frac{1}{k \gamma^0 - (\gamma^0)^2 } \right] ,
\label{acheTM}
\end{equation}
where, after rotation $\omega \rightarrow i \xi$,  $\gamma^0 =\sqrt{k^2+\xi^2/c^2}$ and
$\eta_T=\sqrt{k^2 + \epsilon^{\perp}(k,i\xi) \xi^2/c^2}$ from
(\ref{rTE}) and (\ref{transversal}). Equating $(H^{\rm TM} - 1)/(H^{\rm TM}+1)$ to the expression (\ref{rTM}) for the TM reflection coefficient, one obtains $H^{\rm TM}(k,i\xi)=\overline{\epsilon}(i\xi) \gamma^0 / \chi$. Using this expression in (\ref{acheTM}) one can derive
the longitudinal permittivity 
\begin{equation}
\epsilon^{\|}(k,i\xi) = \frac{k}{\gamma^0} 
\times \left[
\frac{\overline{\epsilon}(i\xi) \gamma^0}{\chi} - 1 + \frac{k}{\gamma^0} + \frac{\xi^2/c^2}{\gamma^0} 
\left( \frac{1}{\eta_T} - \frac{1}{\gamma^0} \right) +
\frac{k \xi^2/c^2}{\gamma^0} \left( \frac{1}{k \eta_T - \eta_T^2} - \frac{1}{k \gamma^0 - (\gamma^0)^2} \right) \right]^{-1}.
\end{equation}
As follows from the above considerations, our theory for Casimir
forces with media with low density of charge carriers can be directly
connected to spatial dispersion, as mentioned by us in \cite{us}. 


\section{Conclusions}

In this paper we have expanded on our previous work \cite{us}
to compute Casimir-Lifshitz forces between 
bodies with low density of charge carriers (intrinsic
semiconductors, dielectrics, disordered systems, etc) 
taking into account
Debye-H\"uckel screening and charge drift. 
Our approach is based on the classical Boltzmann
transport equation, and is applicable to non-degenerate systems with an energy gap. We have shown how
the finite conductivity of such systems modifies the Casimir-Lifshitz force between such materials, and made numerical predictions for the force using germanium and silicon plates.
Our theory can be seen as a special case of spatial dispersion, and provides a simple way to take into 
account nonlocal effects in terms of readily available
material properties. We have explicitly shown that
our theory is compatible with Nernst theorem of thermodynamics. This is in agreement with previous work \cite{Sernelius} that demonstrated that spatial dispersion resolves the issues with 
Nernst theorem. Work related to our approach, totally based on nonlocal dielectric responses, recently appeared \cite{Svetovoy}, extending our analysis to degenerate systems.


We acknowledge correspondence with C. Henkel, F. Intravaia, L.P. Pitaveskii, and discussions with S.A. Ellingsen, G.L. Klimchitskaya, V.M. Mostepanenko, and F.S.S. Rosa. D.A.R.D. is grateful to Victor Dodonov and the other
organizers of the workshop on ``60 Years of the Casimir Effect".
He also acknowledges the support of the U.S. Department of Energy through
the LANL/LDRD program for this work.


\begin{thebibliography}{100}

\bibitem{Lifshitz} Lifshitz E M 1956 {\it Zh. Eksp. Teor. Fiz} {\bf 29}, 94 [{\it Sov. Phys. JETP} {\bf 2} 73]

\bibitem{Pitaevskii}
Pitaevskii L P 2008 {\it Phys. Rev. Lett.} {\bf 101} 163202

\bibitem{us} Dalvit D A R and Lamoreaux S K 2008 {\it Phys. Rev. Lett.} {\bf 101} 163203

\bibitem{screening} Landau L D and Lifshitz E M 1980 {\it Statistical
Physics, Part 1} (Pergamon Press, Oxford)

\bibitem{Sumi} Sumi M 1967 {\it Japanese Journal of Applied Physics}
{\bf 6} 688

\bibitem{Thiennot} Thiennot J 1972 {\it Le Journal de Physique} {\bf 33} 219

\bibitem{Frey} Frey B J, Levinton D B, and Madison T J 2006
{\it Proceedings of SPIE} {\bf 6273 II}; arXiv:physics/0606168

\bibitem{parameters} http://www.ioffe.ru/SVA/NSM

\bibitem{relaxation} http://www.iue.tuwien.ac.at/phd/palankovski/node51.html

\bibitem{controversy} See, for example, Bezerra V B, Klimchitskaya G L,  Mostepanenko V M  and Romero C 2004 {\it Phys. Rev. A} {\bf 69}, 022119; H{\o}ye J S, Brevik I, Ellingsen S A and Aarseth J B 2007 {\it Phys. Rev. E}
{\bf 75} 051127, and references therein.

\bibitem{Rumer} See, for example, Rumer Yu B and Ryvkin M Sh 1980 
{\it Thermodynamics, Statistical Physics, and Kinetics}
(Mir Publishers, Moscow)

\bibitem{correctNernst}
A more precise definition of the Nernst theorem is that in the thermodynamic limit ($N \rightarrow \infty$)
the entropy per particle $S/N=k_{\rm B} \ln(\Omega_N)/N$
should verify $\lim_{N \rightarrow \infty} \ln(\Omega_N) /N
\rightarrow 0$ at $T=0$, limiting the possible number
of ground states for a many-body system (see, for
example, Kardar M 2007 {\it Statistical
Physics of Particles} (Cambridge University Press, Cambridge)). Note that as long as the ground-state degeneracy does not grow exponentially with the system size, then the entropy
per particle does vanish at zero temperature in the thermodynamic limit. 

\bibitem{Intravaia} Intravaia F and Henkel C 2008 {\it J. Phys. A: Math. Theor.} {\bf 41} 164018 

\bibitem{MMlongPRA} See, for example, Rosa F S S, Dalvit D A R, and Milonni P W 2008 {\it Phys. Rev. A} {\bf 78} 032117

\bibitem{Barash} Ginzburg V L and Barash Yu S 1975
{\it Sov. Phys. Ups.} {\bf 18} 931

\bibitem{Sernelius} Sernelius B E 2005 {\it Phys. Rev. B} {\bf 71} 235114; 2006 {\it J. Phys. A:Math. Gen.} {\bf 39} 6741

\bibitem{Esquivel} Esquivel-Sirvent R, Villarreal C, Moch\'an W L, Contreras-Reyes A M,
and Svetovoy V B 2006 {\it J. Phys. A:Math. Gen.} {\bf 39} 6323

\bibitem{Kliewer} Kliewer K L and Fuchs R 1969 {\it Phys. Rev.} {\bf 181} 552

\bibitem{Svetovoy} Svetovoy V B 2008 {\it Phys. Rev. Lett.} {\bf 101} 163603

\end{thebibliography}
\end{document}